\documentclass{article}
\usepackage{ijcai22}

\usepackage{tikz}
\graphicspath{ {./Figures/} }

\usepackage{times}
\usepackage{soul}
\usepackage{url}
\usepackage[hidelinks]{hyperref}
\usepackage[utf8]{inputenc}
\usepackage[small]{caption}
\usepackage{graphicx}
\usepackage{amsmath}
\usepackage{amsthm}
\usepackage{booktabs}
\usepackage{algorithm}
\usepackage{algorithmic}
\usepackage[T1]{fontenc}    
\usepackage{amsfonts}       
\usepackage{nicefrac}       
\usepackage{microtype}      
\usepackage{subcaption}

\newcommand{\bit}{\{0,1\}}
\newcommand{\ignore}[1]{}
\newcommand{\PFL}{{ \Pi_{\sf PPFL}}}
\newcommand{\PDH}{{ \Pi_{\sf DH}}}
\newcommand{\secparam}{{\lambda}}
\newcommand{\Setup}{{\sf Setup}}

\newtheorem{myprot}{Protocol}

\usepackage{tabularx}
\newcounter{protocol}
\newenvironment{protocol}[1]
  {\par\addvspace{\topsep}
   \noindent
   \tabularx{\linewidth}{@{} X @{}}
    \\\hline
    \refstepcounter{protocol}\textbf{Protocol \theprotocol} #1 
    \\\hline}
  {\hrule
  }
\usepackage{enumitem}

\usepackage{amsthm}
\usepackage{amsmath}
\usepackage{amsfonts}
\newtheorem{theorem}{Theorem}

\theoremstyle{definition}
\newtheorem{definition}{Definition}

\title{Collusion Resistant Federated Learning with\\Oblivious Distributed Differential Privacy}

\author{
David Byrd$^1$
\and
Vaikkunth Mugunthan$^2$
\and
Antigoni Polychroniadou$^3$
\And
Tucker Hybinette Balch$^3$\\
\affiliations
$^1$Bowdoin College\\
$^2$Massachusetts Institute of Technology\\
$^3$J.P. Morgan AI Research\\
\emails
d.byrd@bowdoin.edu, 
vaik@mit.edu, 
\{antigoni.o.polychroniadou, tucker.balch\}@jpmchase.com
}

\begin{document}

\maketitle

\begin{abstract}
Privacy-preserving federated learning enables a population of distributed clients to jointly learn a shared model while keeping client training data private, even from an untrusted server.
Prior works do not provide efficient solutions that protect against collusion attacks in which parties collaborate to expose an honest client's model parameters.
We present an efficient mechanism based on oblivious distributed differential privacy that is the first to protect against such client collusion, including the ``Sybil'' attack in which a server preferentially selects compromised devices or simulates fake devices.
We leverage the novel privacy mechanism to construct a secure federated learning protocol and prove the security of that protocol.  We conclude with empirical analysis of the protocol's execution speed, learning accuracy, and privacy performance on two data sets within a realistic simulation of 5,000 distributed network clients.
\end{abstract}

\section{Introduction}

Modern practitioners of machine learning often need to train models from large data sets distributed across many devices.  In the past, such data would be centralized for analysis, but the practice has given rise to serious concerns around user privacy, lack of permission to transfer data, and far-reaching consequences when centralized data stores are breached.

Federated learning is a recent technique that addresses these concerns by training on each local data segment individually, then transmitting and combining only the resulting model parameters. \cite{bonawitz2017practical,kairouz2019advances}  A typical approach is:  The trusted server selects some users to train a new model on their local data, starting from the most recent shared model.  Each user sends their local model weights to the server, which computes an average-weight shared model.  The new shared model is sent to all users, and the process repeats.  In some cases, however, individual user privacy can still be compromised by using the trained model to infer certain details of the training data set.  \cite{shokri2017membership,nasr2018comprehensive}.

Two key approaches have been proposed to address this problem.  The first is \emph{differential privacy}, which perturbs values to guarantee statistical indistinguishability for individual inputs.  \cite{dwork2006our}  This can be applied to federated learning by having each client modify its local model weights by adding randomly-generated values (potentially reducing model accuracy), with the result that a party obtaining the transmitted weights will still have uncertainty over the original weights. 
The second approach, which does not compromise accuracy, is \emph{secure multi-party computation} (MPC).  \cite{GMW87}  An MPC protocol, which allows parties to collaboratively compute a common function of interest without revealing their private inputs, is considered secure if the parties learn the computational output and nothing else.

We build on a recent line of research that combines differential privacy and MPC to produce a secure federated learning protocol.  \cite{bonawitz2017practical,jayaraman2018distributed}   These prior works provide strong protection against undesired inference by the server, but the collusion of enough clients can reveal the noisy weights of an honest client, and the scale of that noise is limited by the need for an accurate model.

We propose a novel, efficient mechanism that protects against any attempt to undermine differential privacy by collusion of $n-1$ out of $n$ total clients.  Unlike prior works, we offer a protocol where the noise for each party is added in an oblivious way.  Obliviousness \emph{can} be achieved by running the noise generation inside the MPC, but such solutions are based on heavy cryptography machinery involving a significant amount of public key operations or incur increased communication complexity. \cite{jayaraman2018distributed,champion2019securely} In this work we focus on the concretely efficient aggregation protocol of Bonawitz et al. without drop-out parties which does not involve any public key operations in the learning phase. \cite{bonawitz2017practical} We therefore provide the first practical protection against $n-1$ attacks by constructing an efficient oblivious distributed differentially private aggregation protocol.

\section{Background}

\subsection{Secure Multiparty Computation}
\label{sec:overview-MPC}

Consider $n$ parties $P_1,\ldots,P_n$ that hold private inputs $x_1 ,\ldots ,x_n$ and wish to compute some arbitrary function $(y_1,\ldots,y_n) = f(x_1,\ldots,x_n)$, where the output of $P_i$ is $y_i$. Secure Multi-Party Computation (MPC) enables the parties to compute the function using an interactive protocol such that each party $P_i$ learns exactly $y_i$ and nothing else.  \cite{GMW87}  (See Appendix A.1 for further detail.)

\subsection{Differential Privacy}
\label{sec:overview-DP}

Differential privacy states that if there are two databases that differ by only one element, they are statistically indistinguishable from each other. In this work we use the Laplacian mechanism which preserves $\epsilon$-differential privacy \cite{dwork2006our}. (See Appendices A.2, A.3, A.4 for further detail.)

\begin{definition}($\epsilon$-differential privacy  \cite{dwork2006calibrating})
A randomized mechanism $\mathcal{A}$ preserves $\epsilon$-differential privacy ($\epsilon$-DP) if for any two neighboring datasets ${D}_1,{D}_2$ that differ by one element, and for all subsets of possible answers $\mathcal{S} \subseteq Range(\mathcal{A})$,~~$ \text{Pr [}\mathcal{A}(D_1)\in \mathcal{S} \text{]}\leq e^\epsilon \text{ Pr [}\mathcal{A}(D_2)\in \mathcal{S}\text{]}$.
\end{definition}

\subsection{Federated Logistic Regression Classifiers}\label{sec:log}

Logistic regression is a machine learning algorithm used to solve the problem of binary linear classification. Assume one of $n$ parties is called $P_i$ and has a local data set consisting of instances $x^{(i)}=(x_1^{(i)}, x_2^{(i)},...., x_m^{(i)})$, where $m$ is the number of features, and their corresponding labels $y^{(i)}$.

Party $P_i$ uses its training examples $(x^{(i)},y^{(i)})$ to learn a logistic classifier with weights $w_{i}$. The weights are obtained by solving the following optimization problem where $f(x_k^{(i)})=w^Tx_k^{(i)}$ and $t_i$ is the number of training examples of $P_i$:
\begin{equation}
    w_{i}= \underset{w}{\arg\min}\frac{1}{t_i}\sum_{k=1}^{t_i} log (1+e^{-y_k^{(i)}f(x_k^{(i)})})
\end{equation}
In order to minimize the loss function, we make use of gradient descent, an iterative optimization algorithm, calculating the optimal $w$ iteratively as 
$w^{j+1} \leftarrow w^j - \alpha \nabla L(w^j)$, where $\alpha$ is the learning rate, $j$ is the iteration, $w^0=0$, and $\nabla L$ is the gradient of the loss function.  Our local logistic regression is a vector-based re-implementation of Jayaraman et al.~\cite{jayaraman2018distributed}.

Privacy-preserving federated learning allows a large number of parties to learn a model while keeping their local training data private.  Parties first train local models on their local data and coordinate with a server to obtain a global model.  Given $n$ parties, let  ${w_i}$, for $i \in 1$ to $n$, represent the local model estimator after minimizing the objective function.  Then ${W}=\frac{1}{n}\sum_{i=1}^{n}{w_i} + \eta$, where $\eta$ is the differentially private noise added to the cumulative model.

\mathchardef\mhyphen="2D
According to Jayaraman et al., for $1$-$Lipschitz$ the global sensitivity for a multi-party setting is $\frac{2}{n*k*\alpha}$, where k is the size of the smallest dataset amongst the $n$ parties, and $\alpha$ is the regularization parameter. \cite{jayaraman2018distributed} Hence, $\eta=\mathcal{L}(\frac{2}{n*k*\alpha*\epsilon})$, where $\epsilon$ is the privacy loss parameter.
In our protocol, each client will add noise to the weights of the trained local model.

\subsection{Network Topology \& Threat Model}
\label{sec:net}

As is common in the federated learning setting, we opt for a star network topology, where there is one central party that is connected to all other parties. This central server can be distinct from the $n$ original parties.

The protocols that we describe and compare against are secure in the semi-honest model. A semi-honest adversary follows the protocol correctly but tries to learn as much as possible about the inputs of the uncorrupted parties from the messages it receives. Furthermore, if there are multiple semi-honest corruptions, we allow the adversary to combine the views of the corrupted parties to potentially learn more information.  See Appendices B.1 and B.3 for security in the malicious model where the corrupted parties misbehave, and Appendix B.5 for communication protocol diagrams.

\section{Approach}

Our approach combines secure multi-party aggregation with oblivious distributed differential privacy to better secure federated learning against $n-1$ collusion attacks.  In this work, we consider logistic regression as the local learning method, and each client update includes the weights of that logistic regression.  The server receives the weights from all clients at each iteration and computes a new global model using the average of the client updates for each weight.  Recall from the Introduction the literature demonstrating that private client data can be inferred from the trained model weights, which is clearly undesirable.  The general task, then, is to secure each client's locally trained model weights against discovery while still learning an accurate shared model.  We note that the collusion problem can be solved using generic MPC, but such generic solutions are impractical due to computational inefficiency.  Our contribution is a practical and efficient solution to this problem using lightweight cryptographic tools.

\subsection{Eliminating weight leakage} We use a secure weighted average protocol running across $n$ clients to hide each client's model weights from the server where each weigh is sent to the server encrypted/masked. The underlying secure aggregation protocol for online/non- drop-out clients we use appeared in the work of Bonawitz et al.~\cite{bonawitz2017practical}, in which clients send individual updates to the server in an encrypted manner.


\begin{table*}[h!]
\begin{protocol}{Privacy-Preserving Federated Logistic Regression Protocol $\PFL$ for a single weight\\}
\\The protocol $\PFL$ runs with parties $P_1,\ldots, P_n$ and a server $S$. It proceeds as follows:

\textbf{Inputs:} For $i \in [n]$, party~$P_i$ holds input dataset~$D_i$. \\

{\bf \boldmath $\PFL.\Setup(1^\secparam)$}: Each party $P_i$ for $i\in[n]$ proceeds as follows for all $j\in[n]$ ($i\neq j$):

\begin{itemize}
\item Generate random variables $\gamma_{i,j}^b$ and $\bar{\gamma}_{i,j}^b$ for $b\in\bit $ from the gamma $\mathcal{G}(1/n,scale)$ distribution with $scale= 2/(n*len(D_i)*\alpha*\epsilon)$. See Section \ref{sec:log} for the details on $scale$. 

\item Generate random masks $s_{i,j} \in \mathbb{Z}_q$.\smallskip
\item Compute masked noises $\eta_{i,j}^0=s_{i,j}+\gamma_{i,j}^0 - \bar\gamma_{i,j}^0$ and  $\eta_{i,j}^1=s_{i,j}+\gamma_{i,j}^1 - \bar\gamma_{i,j}^1$. 
\item Run the Diffie-Hellman key Exchange protocol $\PDH$ to obtain a common shared key $r_{i,j}$ with each party $P_j$ which is only known between parties $P_i$ and $P_j$ ($r_{i,j}$ is not known to $S$).

\item Each party $P_i$ sends $\eta_{i,j}^0,\eta_{i,j}^1$ to $S$ who permutes and randomizes them and forwards to party $P_j$. 

\end{itemize}

Given the above setup, we can compute the federated logistic regression model as follows: 

\smallskip
{\bf \boldmath $\PFL.{\sf WeightedAverage}(D_i,\{r_{i,j}\}_{j\in[n]})$}:

\begin{enumerate}
\item []\textbf{Round 1:} Each party $P_i$ proceeds as follows: 
\begin{itemize}
    \item Compute the weights $w_{i}$, using Equation (1), of the local logistic classifier obtained by implementing regularized logistic regression on input $D_i$.

We describe the algorithm for a single weight, denoted by $w_i$:

\item Generate a random bit vector $b=(b_1,\ldots,b_n)$ and send $y_i$ to the server $S$:
$y_i := w_i + \sum_{j=i+1}^n r_{i,j} - \sum_{k=1}^{i-1}r_{k,i} + \sum_{j=1}^{n }\eta_{j,i}^{b_j}- \sum_{j=1}^{n}s_{i,j}   \bmod p\ .$
\end{itemize}

\item []\textbf{Round 2:} The server computes $W=\big(\sum_{i=1}^{n}y_i\bmod p \big)/n  $ and sends W to all parties. 
\end{enumerate}

{\bf \boldmath $\PFL.Output(1^\secparam,W)$}: Each party $P_i$ upon receiving $W$ repeats $\mathsf{WeightedAverage}$ for the next weight or iteration of the logistic regression with locally updated common keys $r_{i,j}'$. (See Section C in the Appendix for further discussion.)

\end{protocol}
\end{table*}

\subsection{Eliminating weighted average leakage} Using the secure aggregation protocol of ~\cite{bonawitz2017practical} hides all information about client weights from the server, but the final shared model can still reveal information about individual client weights and subsequently a client's local data set. Given the output which is the average of each model weight, $n-1$ clients working together can remove their weights to discover the exact model weights of the remaining ``honest'' client.

Previous approaches augment the secure aggregation protocol with differential privacy to mitigate the impact of client data exposure.  Under these protocols, each client independently generates and adds random noise to each model weight prior to transmission, so even in the case of $n-1$ client collusion, only ``noisy'' weights can be recovered.  This is a definite improvement, but unlike MPC it is a lossy one, and the scale of the added noise is limited by a trade-off against model accuracy.

We introduce a novel and efficient \emph{oblivious distributed} differentially private mechanism.  In prior works, each client picks its own local noise. By contrast, we offer a protocol where the noise for each party is added in an oblivious way. More specifically: For each weight, each client receives a tuple of encrypted noise terms from each other client and adds only a subset of them. Thus, a party $P$ does not know the cleartext noise added to its weight and the other parties do not know which noise term is chosen by $P$.

We show that the information leakage on the honest client's weights after the collusion attack is smaller than previous approaches.  Our task is to enable the parties to calculate the sum of their inputs (i.e., $W=\sum_{i=1}^n w_i$), while ensuring privacy for an honest party in the presence of a collusion attack given $W$. In prior works if $n-1$ parties collude, they subtract their weights and noise terms from $W$ and then the final noise remaining in the transmitted weight of the honest party $w_h$ is a single value chosen by the honest party. In our case, if $n-1$ parties collaborate then the final noise remaining in the transmitted weight of the honest party $w_h$ is $n-1$ times larger, because the corrupted parties cannot subtract their noise terms. 

At a high level, in our scheme, each client sends two encrypted noisy terms (permuted and randomized by the server) per model weight to the other clients, but each receiving client chooses only one of the two to add to each weight. Thus even if parties collude they cannot subtract a significant number of noise terms since they do not know which noise terms the honest client chose.

\section{Secure Weighted Average Protocol}
\label{sec:secure-aggr}

\subsection{Our Protocol}
\label{sec:secure-aggr-prot}

We formally describe our weighted average protocol $\PFL$, depicted in Protocol 1, for secure logistic regression performed by a set of clients $(P_1,\ldots,P_n)$ and a server $S$.  

During setup, every pair of parties $P_i$ and $P_j$ will share some common randomness $r_{i,j}=r_{j,i}$. In the online weighted average phase, client $P_i$ sends its weights masked with these common random strings, adding all $r_{i,j}$ for $j > i$ and subtracting all $r_{i,k}$ for $k < i$.  That is, $P_i$ sends to server $S$ the following message for its data $w_i$: ${y}_i := (w_i + \sum_{j = i+1}^n r_{ij} - \sum_{k =1}^{i-1}r_{ki}) \bmod p$. Each weight received by the server is masked by $n-1$ large random numbers $r$, so it cannot accurately reconstruct any client's true model weight $w_i$.  Because the total randomness applied to the weights sums to zero once the server computes $W$ in round $2$ of $\PFL$, the averaged final model $W$ will be identical to one calculated without security.

To establish common randomness $r$, each pair of parties run the standard Diffie-Hellman Key exchange protocol from the literature \cite{DH76} communicating via the server.  (See Appendix D.5 for description and listing.)

The protocol is given for a single iteration of federated logistic regression.  For a detailed explanation of how the parties \emph{locally} update their $r$ masks to be used in the next weight and next iteration of the protocol, see Section C of the Appendix. We generate the Laplacian noise in a distributed way by the use of gamma distributions $\mathcal{G}$ given that the Laplace distribution $\mathcal{L}$ can be constructed as the sum of differences of i.i.d. gamma distributions.
To run machine learning algorithms and the DP mechanism which computes on rational values,
we use field elements in a finite field $\mathbb{Z}_q$ to represent
the \emph{fixed-point values}. Concretely, for a fixed-point value $\bar{x}$ 
with $k$ bits in the integer part and $f$ bits in the decimal part, we
use the field element $x:=2^f\cdot\bar{x} \mod q$ in $\mathbb{Z}_q$ to
represent it.

\begin{figure*}[ht]
\begin{minipage}[b]{0.33\linewidth}
\includegraphics[width=\linewidth]{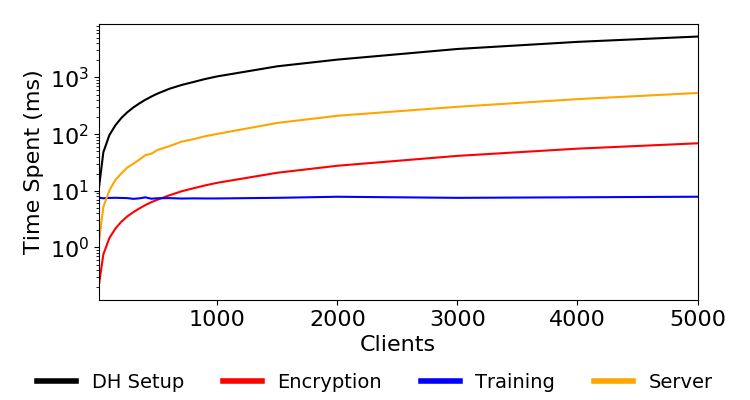}
\caption{Component timing on Graph 1.}
\label{fig:component_time}
\end{minipage}
\hfill
\begin{minipage}[b]{0.33\linewidth}
\includegraphics[width=\linewidth]{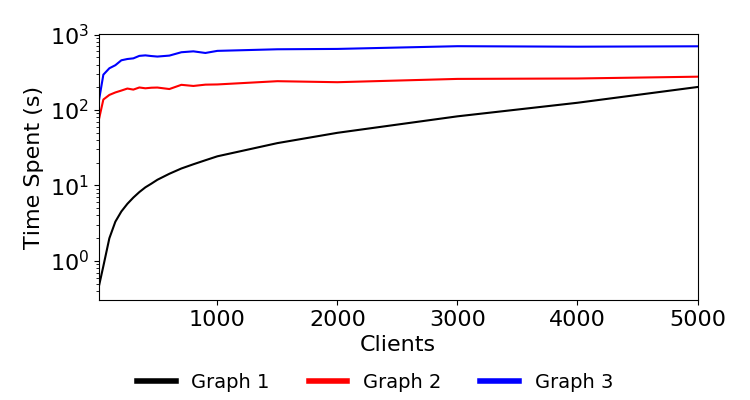}
\caption{Total protocol time for Graphs 1-3.}
\label{fig:total_time}
\end{minipage}
\hfill
\begin{minipage}[b]{0.33\linewidth}
\includegraphics[width=\linewidth]{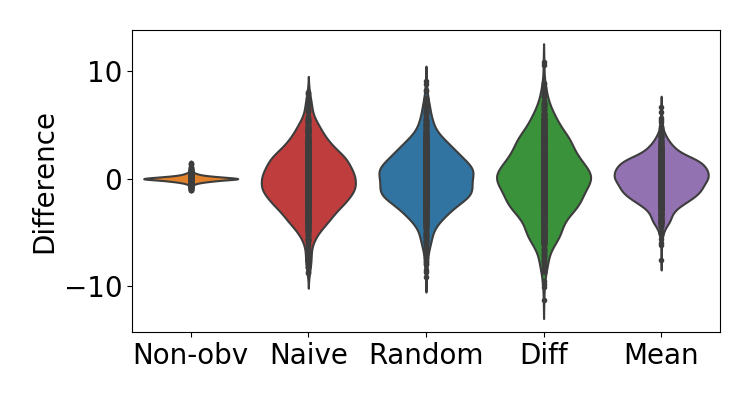}
\caption{Error in attacker weight estimates.}
\label{fig:violin}
\end{minipage}%
\end{figure*}

\begin{figure*}[ht]\hfill
\begin{subfigure}{0.4\textwidth}

\centering
\includegraphics[width=\textwidth]{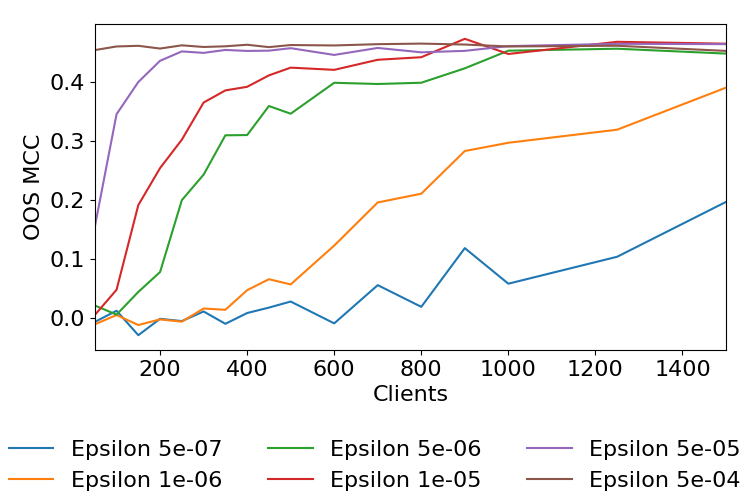}
\caption{Final shared model by $\epsilon$ privacy loss parameter.}
\label{fig:accuracy_by_parties}

\end{subfigure}\hfill
\begin{subfigure}{0.4\textwidth}

\centering
\includegraphics[width=\textwidth]{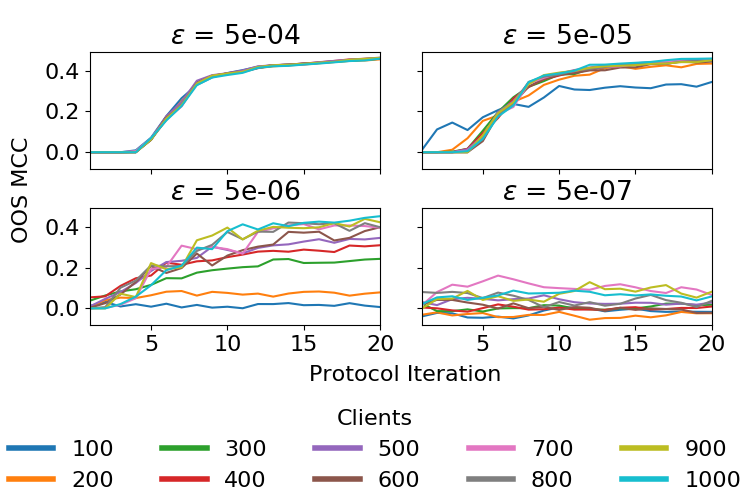}
\caption{Per protocol iteration by $\epsilon$ and client count.}
\label{fig:neurips_training}

\end{subfigure}\hfill~
\caption{Out of sample performance (Matthews Correlation Coefficient) for our oblivious distributed differential privacy protocol.}
\end{figure*}

\subsection{Security of our Protocol}
\label{sec:secure-aggr-prot}

We prove that
our protocol protects the privacy of honest users 
in the semi-honest setting given the topology in Section \ref{sec:net}. In particular we show the following theorem.

\begin{theorem}
    Suppose $n$ clients $P_1,\dots,P_n$ each hold private input $w_i$, and they wish to rely on a server $S$ to compute the sum $f(w_1,\dots,w_n)=\sum_iw_i$. There exists a protocol $\PFL$, returning the sum which does not leak any information about
the other clients' inputs
except what can be inferred
from the sum and offers collusion-privacy against a coalition of up to $t\leq n-1$ clients. 
\end{theorem}

In Appendix B.1, we further formalize and prove our theorem, and consider security against $t\leq n-1$ semi-honest clients and a curious server, and against $t$ malicious users.

Next we argue that the error term on the honest client’s inputs after the collusion attack of $t$ parties is larger than previous approaches.  For this, we require the following additional property. Consider the case of $n-1$ collusion; we define collusion privacy as follows:

\emph{\bf Collusion-Privacy}: An $n$-party protocol provides \emph{Collusion-Privacy}, for an aggregation function $f$ and a probability distribution $\mathcal{D}$, if any adversary, who controls all parties except client $P_h$, learns no more than the honest party’s values $w_h + \eta$ where $\eta\leftarrow \mathcal{D}$ and $f(w_1,\dots, w_n)$.

In prior works if $n-1$ parties collude then the final noise left in the weight of the honest party $w_h$ is a single value from $\mathcal{D}$. In our case, if $n-1$ parties collaborate then the final noise left in the weight of the honest party $w_h$ is $n-1$ times larger than $\mathcal{D}$ since the corrupted parties cannot subtract their exact noise terms.

To measure the error, we quantify the difference between $f(D)$ and its perturbed value $\hat{f}(D)$ which is the error introduced by the differential private mechanism of the secure aggregation protocol.

\begin{definition}
(Error function) Let $D \in \mathcal{D}$, $f: \mathcal{D} \rightarrow \mathbb{R}$, and let $\delta = \frac{|f(D)-\hat{f}(D)|}{|f(D)|+1}$ (i.e., the value of the error). The error function is defined as $\mu= \mathbb{E}(\delta)$. The expectation is taken on the randomness of $\hat{f}(D)$. The standard deviation of the error is $\sigma = \sqrt{Var(\delta)}$.
\end{definition}

After the execution of Protocol 1, parties receive the noisy sum of their inputs, i.e., $W=\sum_{i=1}^n w_i$, 
In prior non-oblivious works if $n-1$ parties collaborate and remove their weights from $w$ then the final noise added to the weight of the honest party $w_h$ is a value from $\mathcal{L}(\lambda)$, and hence, the error is $\mu = \frac{1}{|W|+1} \mathbb{E}|\mathcal{L}(\lambda)| = \frac{\lambda}{|W|+1}$.

In our oblivious case, if $n-1$ parties collaborate then the final noise added to the weight of the honest party $w_h$ is $n-1$ times larger than $\mathcal{L}(\lambda)$, and hence, the error is $\mu = \frac{1}{|W|+1} \mathbb{E}|\sum_{i=1}^{n-1} \mathcal{L}(\lambda)| = \frac{(n-1)\cdot \lambda}{|W|+1}$. 

However, in practice even if the parties cannot subtract their exact noise terms they can still try to subtract the average of the noise terms, or one of the two noise terms, or use the leakage $L$ to reduce the amount of error. In our protocol we consider the leakage $L$ learned from the difference of the noise terms $\eta^0,\eta^1$. Note that this leakage does not affect the error function given later in Definition 2. In Section \ref{sec:attack} and Figure \ref{fig:collusion_dense} we empirically show that such an attack is little better than the attack of subtracting nothing.

Note that the output of the aggregation protocol, $W + \eta$, is generated such that $\eta$ follows exactly the same distribution in both non-oblivious and oblivious cases, but the noise left after an $n-1$ attack against the oblivious case is higher. For further discussion, see Appendix B.2 on collusion privacy.

\section{Experiments}

We empirically evaluated our protocol using ABIDES, an open source simulation platform originally designed for financial markets \cite{byrd2019abides} and later adapted for federated learning \cite{byrd2020differentially}.

Following the agent-based approach of these prior works, we simulated our oblivious protocol for 5,000 distributed clients and analyzed the timing, accuracy, and privacy of the empirical results.

\subsection{Experimental Dataset and Method}
We evaluated our protocol's performance using the Adult Census Income dataset \cite{repository_ml}, which provides 14 input features such as age, marital status, and occupation, that can be used to predict a categorical output variable identifying whether (\emph{True}) or not (\emph{False}) an individual earns over \$50K USD per year.  We used a preprocessed version of the dataset from Jayaraman et al. following the method of Chadhuri et al. which transformed each categorical variable into a series of binary features, then normalized both features and examples, resulting in 104 features for consideration. \cite{jayaraman2018distributed,chaudhuri2011differentially}  We added a constant intercept feature to permit greater flexibility in the regression.  Of the 45,222 records in our cleaned data set, there were 11,208 positive examples (about 25\%), representing a moderately unbalanced dataset.

The dataset was loaded only once per complete simulation of the protocol, after which a randomized train-test split (75\% vs 25\%) was taken.  Once per round of federated learning, each client randomly selected 200 rows from the training data as its ``local'' data.  The holdout test data was the same for all clients, and no client ever trained on it.  All clients implemented Protocol 1 as previously described. 

\begin{figure*}[ht]
\centering
\includegraphics[width=0.8\textwidth]{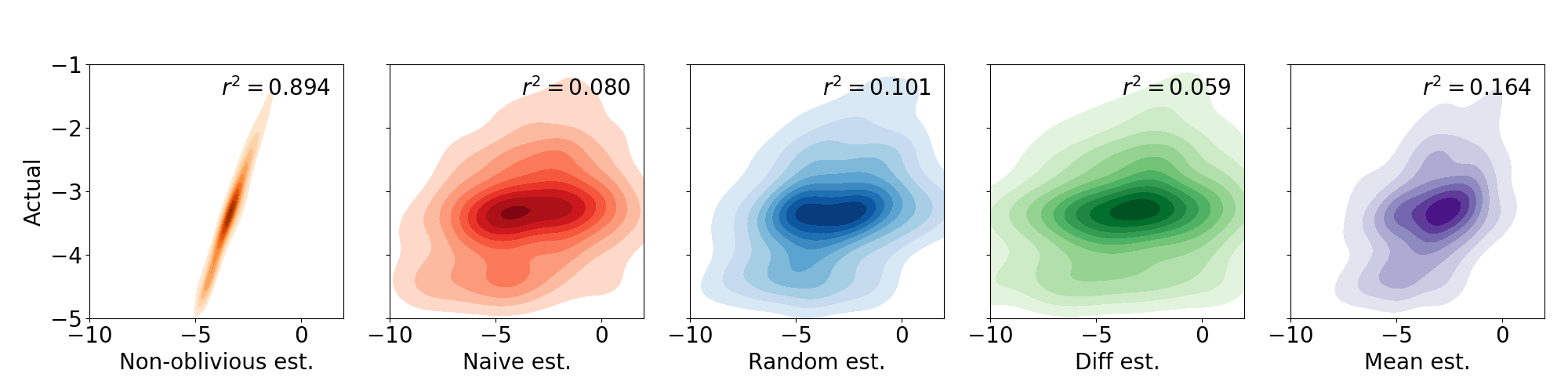}
\caption{Density plot of actual versus estimated honest party weight over 1,000 iterations of the $n-1$ collusion attack.}
\label{fig:collusion_dense}
\end{figure*}

\subsection{Protocol Timing Results}

The ABIDES simulation permits construction of an arbitrary network graph with defined pairwise connectivity, minimum latency, and parameters for randomly selected ``jitter'' with nanosecond resolution.  It captures the real elapsed runtime of each client activity and appropriately delays both sent messages and the earliest time at which a client may act again.  Using these features, we have estimated the temporal load of Protocol 1.  The mean time required to run the protocol simulation on a single Intel Xeon X5650 CPU core (2.6GHz) ranged from 32 seconds to 12 hours for 100 to 5,000 parties.

Figure \ref{fig:component_time} summarizes the time spent performing each section of our protocol on the adult census income data set: \textbf{Diffie-Hellman Setup} one time per client, \textbf{Encryption} of the weights and local model, \textbf{Training} per client per protocol iteration, and \textbf{Server} aggregation time per protocol iteration.  Figure \ref{fig:total_time} shows the estimated time required to run the full protocol (not the simulation) for three different network graphs: \textbf{Graph 1} places all participants around New York City, \textbf{Graph 2} places the server in New York City and clients around London, \textbf{Graph 3} places the server in New York City and clients all over the world.  Tabular data is presented in Appendix D.2.

All experiments comprised 20 rounds of secure federated learning, with each client running 50 iterations of local regression training at each round.  Latency is the most significant time component for small participant networks, but as the population size grows, computation effort surpasses it.  Fortunately, the two largest components of computational time growth represent work performed only once per client for the entire protocol, and work performed only by the server.

\subsection{Protocol Accuracy Results}

The secure multi-party aggregation component of Protocol 1 is lossless, because the MPC encryption elements sum to zero in each shared model.  Differential privacy introduces shared model accuracy loss inversely proportional to the $\epsilon$ privacy loss parameter.  Smaller selections of $\epsilon$ result in more uncertainty about a client's local weights when other parties collude to reveal them, but increasingly confound learning.  For example, in our protocol experiments with 200 clients, final model accuracy worsens dramatically once $\epsilon < 5e-5$.

{\bf Matthews Correlation Coefficient:} Because of the significant (3:1) class imbalance in our data, we assess accuracy using the Matthews Correlation Coefficient (MCC) \cite{matthews1975comparison}, a contingency method of calculating the Pearson product-moment correlation coefficient (with the same interpretation), that is appropriate for imbalanced classification problems. \cite{baldi2000assessing,pearson1895note,powers2011evaluation,evans1996straightforward} (See Appendix D.1 for more detail.)

In Figure \ref{fig:accuracy_by_parties} we show the MCC of our protocol's final shared model predictions against the correct values for a range of $\epsilon$.  As expected, smaller $\epsilon$ harms the accuracy of the learned model.  Thus there is a dynamic lower bound, varying with population size, on useful values of $\epsilon$.  For example when considering out of sample $MCC(n)$, with $n$ being the client population size, in our experiments with $\epsilon=1e-5$: $MCC(100)=0.005$, $MCC(200)=0.254$, and $MCC(500)=0.423$.  For all evaluated client population sizes (50 to 5,000), models trained under Protocol 1 with $\epsilon \geq 5e-4$ had similar accuracy to unsecured federated learning.  Figure \ref{fig:neurips_training} shows the impact $\epsilon$ can have on each round of federated learning:  with $\epsilon=5e-4$ or $\epsilon=5e-7$, population size does not matter because either all sizes succeed at learning or none do; but with $\epsilon=5e-6$, varying client population sizes learn at vastly different rates.

In our simulated network environment, with 1000 clients implementing the described protocol for federated logistic regression using privacy loss parameter $\epsilon=5e-4$, we found an out of sample final iteration relative accuracy loss (MSE versus learning in the clear) of $1.1e-6$ and an out of sample final iteration relative MCC loss of $0.0018$.

\subsection{Adversarial Data Recovery}\label{sec:attack}
\label{sec-collusion}

Prior works like Bonawitz et al. discuss attacks from a ``snooping'' server which attempts to infer the unencrypted weights of a particular client.  \cite{bonawitz2017practical}  The attacks fail since the server does not have the common random values $r$.

\textbf{Collusion attack:}  We consider the $n-1$ attack, in which \emph{all other clients} conspire to recover the unencrypted model weights of a single ``honest'' client.  Let the honest client be $h$ and the set of $n-1$ colluding clients be $C$.  For a single model weight, let $F = w_h + W_C + T_h + T_C$ be the output of the aggregation protocol at the end of each iteration, where $w_h$ is the honest party's original weight, $T_h$ is the honest party's noise sum, $W_C=\sum_{c\in C}w_c$, $T_C=\sum_{c\in C}T_c$. Note that the sum of the randomness $r,s$ is removed by design from the output when the computation is performed, so MPC cannot defend against this type of attack.

Under prior non-oblivious protocols in which each party generates and adds its own noise locally, the colluding parties know $W_C$ and $T_C$ and can therefore recover the honest party's noisy weights $W_h + T_h$. 
Under our oblivious protocol, the noise which cannot be subtracted in $F$ has a more dispersed distribution that cannot be much narrowed by the colluding parties since $h$ will have received from each colluding party $c$ a choice of two difference of gamma privacy noises $\hat\gamma_{ch}^0$ and $\hat\gamma_{ch}^1$ and $c$ will not know which was selected by $h$. Moreover, since the server permutes and randomizes the encrypted noise terms, $T_C$ is also not precisely known to the colluding parties. For an extreme Sybil attack, the server will have to run a secure shuffle protocol which we have not implemented in this version (see discussion in Appendix B.4).  

We empirically illustrate the dramatic improvement in privacy against an $n-1$ distributed attack by considering five cases.  In each case, the colluding parties attempt to recover $W_h$ and always remove $W_C$.  In the \textbf{Non-Oblivious} case followed by prior works, the corrupted parties can accurately remove $T_C$. In the other four cases, Protocol 1 is attacked, and the corrupted parties must decide how to deal with the unknown noise choices by other parties $i$: under \textbf{Naive} they do nothing additional (i.e., they do not remove any noise terms); under \textbf{Random} each corrupted party $c$ removes either $\hat\gamma_{ci}^0$ or $\hat\gamma_{ci}^1$ at random; and under \textbf{Diff} and \textbf{Mean} each corrupted party $c$ removes the difference or mean of $\hat\gamma_{ci}^0$ and $\hat\gamma_{ci}^1$ respectively.

We consider the recovery attempt across 1,000 full iterations of Protocol 1 with 100 clients participating.  At the end of every iteration, 99 clients  share information in an attempt to recover the unencrypted model weights of the one honest client.  Privacy loss parameter $\epsilon=5e-4$ was selected because it did not cause significant shared model accuracy loss for any tested number of parties.

Figure \ref{fig:collusion_dense} shows a density plot of the honest party's actual model weight versus the collaborators' estimate of that weight.  Figure \ref{fig:violin} summarizes the distribution of the difference between estimated and actual weights for each attack scenario.  The $n-1$ attack is successful ($r^2=0.894$) against the prior non-oblivious protocol, but not successful ($r^2=0.164$ or worse) against our new oblivious protocol.  (For additional discussion of the attacks, see Appendix D.3.)

To further confirm our empirical results, we present a similar attack against the protocol while learning a credit card fraud data set in Appendix D.4 and Figures D.1 and D.2.

\section{Conclusion}
We presented an efficient mechanism for oblivious distributed differential privacy that is the first to secure against collusion attacks on the clients' model parameters, and leveraged that mechanism to construct a secure federated learning protocol.  We also detailed the protocol and proved its security.

To empirically evaluate the protocol in a practical setting, we implemented it for a common data set with 5,000 parties in an open source simulation that has been adapted to the domain of privacy-preserving federated learning, and estimated its accuracy and running time for various client counts and values of the $\epsilon$ privacy loss parameter.  We also conducted an $n-1$ attack and showed that it is effective against prior non-oblivious protocols, but not against our new protocol.

 We have left the case where clients drop off during the protocol as future work.  Our mechanism is therefore well suited to cross-silo federated learning applications where clients are different organizations (e.g. medical or financial) or geodistributed datacenters, as opposed to mobile or IoT devices which can possibly go offline.

\bibliographystyle{named}
\bibliography{oddpfl}

\begin{thebibliography}{}

\bibitem[\protect\citeauthoryear{Baldi \bgroup \em et al.\egroup
  }{2000}]{baldi2000assessing}
Pierre Baldi, S{\o}ren Brunak, Yves Chauvin, Claus~AF Andersen, and Henrik
  Nielsen.
\newblock Assessing the accuracy of prediction algorithms for classification:
  an overview.
\newblock {\em Bioinformatics}, 16(5):412--424, 2000.

\bibitem[\protect\citeauthoryear{Bonawitz \bgroup \em et al.\egroup
  }{2017}]{bonawitz2017practical}
Keith Bonawitz, Vladimir Ivanov, Ben Kreuter, Antonio Marcedone, H~Brendan
  McMahan, Sarvar Patel, Daniel Ramage, Aaron Segal, and Karn Seth.
\newblock Practical secure aggregation for privacy-preserving machine learning.
\newblock In {\em Proceedings of the 2017 ACM SIGSAC Conference on Computer and
  Communications Security}, pages 1175--1191. ACM, 2017.

\bibitem[\protect\citeauthoryear{Byrd and
  Polychroniadou}{2020}]{byrd2020differentially}
David Byrd and Antigoni Polychroniadou.
\newblock Differentially private secure multi-party computation for federated
  learning in financial applications.
\newblock In {\em Proceedings of the 2020 ACM International Conference on AI in
  Finance}, ACM ICAIF '20, New York, NY, USA, 2020. Association for Computing
  Machinery.

\bibitem[\protect\citeauthoryear{Byrd \bgroup \em et al.\egroup
  }{2020}]{byrd2019abides}
David Byrd, Maria Hybinette, and Tucker~Hybinette Balch.
\newblock {ABIDES: Towards} high-fidelity multi-agent market simulation.
\newblock In {\em Proceedings of the 2020 ACM SIGSIM Conference on Principles
  of Advanced Discrete Simulation}, SIGSIM-PADS '20, page 11–22, New York,
  NY, USA, 2020. Association for Computing Machinery.

\bibitem[\protect\citeauthoryear{Champion \bgroup \em et al.\egroup
  }{2019}]{champion2019securely}
Jeffrey Champion, Abhi Shelat, and Jonathan Ullman.
\newblock Securely sampling biased coins with applications to differential
  privacy.
\newblock In {\em Proceedings of the 2019 ACM SIGSAC Conference on Computer and
  Communications Security}, CCS '19, page 603–614, New York, NY, USA, 2019.
  Association for Computing Machinery.

\bibitem[\protect\citeauthoryear{Chaudhuri \bgroup \em et al.\egroup
  }{2011}]{chaudhuri2011differentially}
Kamalika Chaudhuri, Claire Monteleoni, and Anand~D Sarwate.
\newblock Differentially private empirical risk minimization.
\newblock {\em Journal of Machine Learning Research}, 12(Mar):1069--1109, 2011.

\bibitem[\protect\citeauthoryear{Diffie and Hellman}{1976}]{DH76}
Whitfield Diffie and Martin~E. Hellman.
\newblock New directions in cryptography.
\newblock {\em {IEEE} Trans. Information Theory}, 22(6):644--654, 1976.

\bibitem[\protect\citeauthoryear{Dua and Graff}{2017}]{repository_ml}
Dheeru Dua and Casey Graff.
\newblock {UCI} machine learning repository, 2017.

\bibitem[\protect\citeauthoryear{Dwork \bgroup \em et al.\egroup
  }{2006a}]{dwork2006our}
Cynthia Dwork, Krishnaram Kenthapadi, Frank McSherry, Ilya Mironov, and Moni
  Naor.
\newblock Our data, ourselves: Privacy via distributed noise generation.
\newblock In {\em Annual International Conference on the Theory and
  Applications of Cryptographic Techniques}, pages 486--503. Springer, 2006.

\bibitem[\protect\citeauthoryear{Dwork \bgroup \em et al.\egroup
  }{2006b}]{dwork2006calibrating}
Cynthia Dwork, Frank McSherry, Kobbi Nissim, and Adam Smith.
\newblock Calibrating noise to sensitivity in private data analysis.
\newblock In {\em Theory of cryptography conference}, pages 265--284. Springer,
  2006.

\bibitem[\protect\citeauthoryear{Evans}{1996}]{evans1996straightforward}
James~D Evans.
\newblock {\em Straightforward statistics for the behavioral sciences}.
\newblock Brooks/Cole, 1996.

\bibitem[\protect\citeauthoryear{Goldreich \bgroup \em et al.\egroup
  }{1987}]{GMW87}
Oded Goldreich, Silvio Micali, and Avi Wigderson.
\newblock How to play any mental game or {A} completeness theorem for protocols
  with honest majority.
\newblock In {\em Proceedings of the 19th Annual {ACM} Symposium on Theory of
  Computing, 1987, New York, New York, {USA}}, pages 218--229, 1987.

\bibitem[\protect\citeauthoryear{Jayaraman \bgroup \em et al.\egroup
  }{2018}]{jayaraman2018distributed}
Bargav Jayaraman, Lingxiao Wang, David Evans, and Quanquan Gu.
\newblock Distributed learning without distress: Privacy-preserving empirical
  risk minimization.
\newblock In {\em Advances in Neural Information Processing Systems}, pages
  6343--6354, 2018.

\bibitem[\protect\citeauthoryear{Kairouz \bgroup \em et al.\egroup
  }{2019}]{kairouz2019advances}
Peter Kairouz, H~Brendan McMahan, Brendan Avent, Aur{\'e}lien Bellet, Mehdi
  Bennis, Arjun~Nitin Bhagoji, Keith Bonawitz, Zachary Charles, Graham Cormode,
  Rachel Cummings, et~al.
\newblock Advances and open problems in federated learning.
\newblock {\em arXiv preprint arXiv:1912.04977}, 2019.

\bibitem[\protect\citeauthoryear{Matthews}{1975}]{matthews1975comparison}
Brian~W Matthews.
\newblock Comparison of the predicted and observed secondary structure of t4
  phage lysozyme.
\newblock {\em Biochimica et Biophysica Acta (BBA)-Protein Structure},
  405(2):442--451, 1975.

\bibitem[\protect\citeauthoryear{Nasr \bgroup \em et al.\egroup
  }{2018}]{nasr2018comprehensive}
Milad Nasr, Reza Shokri, and Amir Houmansadr.
\newblock Comprehensive privacy analysis of deep learning: Stand-alone and
  federated learning under passive and active white-box inference attacks.
\newblock {\em arXiv preprint arXiv:1812.00910}, 2018.

\bibitem[\protect\citeauthoryear{Pearson}{1895}]{pearson1895note}
Karl Pearson.
\newblock Note on regression and inheritance in the case of two parents.
\newblock {\em Proceedings of the Royal Society of London}, 58:240--242, 1895.

\bibitem[\protect\citeauthoryear{Powers}{2011}]{powers2011evaluation}
David~Martin Powers.
\newblock Evaluation: from precision, recall and f-measure to roc,
  informedness, markedness and correlation.
\newblock {\em Journal of Machine Learning Technologies}, 2011.

\bibitem[\protect\citeauthoryear{Shokri \bgroup \em et al.\egroup
  }{2017}]{shokri2017membership}
Reza Shokri, Marco Stronati, Congzheng Song, and Vitaly Shmatikov.
\newblock Membership inference attacks against machine learning models.
\newblock In {\em 2017 IEEE Symposium on Security and Privacy (SP)}, pages
  3--18. IEEE, 2017.

\end{thebibliography}

\end{document}